\nopagenumbers 
\magnification=\magstep1 
\hsize 6.0 true in 
\hoffset 0.25 true in 
\emergencystretch=0.6 in 
\vfuzz 0.4 in 
\hfuzz 0.4 in 
\vglue 0.1true in 
\mathsurround=2pt 
\def\nl{\noindent} 
\def\nll{\hfil\break\noindent} 
\def\np{\hfil\vfil\break} 
\def\ppl#1{{\leftskip=9cm\noindent #1\smallskip}} 
\def\title#1{\bigskip\noindent\bf #1 ~ \trr\smallskip} 
 
\font\trr=cmr10 
\font\bf=cmbx10 
\font\bmf=cmmib10 
\font\sl=cmsl10 
\font\it=cmti10 
\font\trbig=cmbx10 scaled 1500 
\font\tiny=cmr8 
\def\mb#1{\hbox{\bmf#1}} 
\def\ng{>\kern -9pt|\kern 9pt} 
\def\hi#1#2{$#1$\kern -2pt-#2} 
\def\hy#1#2{#1-\kern -2pt$#2$} 

\def\rtitle{Relativistic massless-boson systems} 
\def\ptitle{The energy of a system of relativistic massless bosons} 
\def\ptitlee{bound by oscillator pair potentials} 

\output={\shipout\vbox{\makeheadline\ifnum\the\pageno>1 {\hrule} \fi 
{\pagebody}\makefootline}\advancepageno} 
 
\headline{\noindent {\ifnum\the\pageno>1 
{\tiny \rtitle\hfil page~\the\pageno}\fi}} 
\footline{} 
\newcount\zz \zz=0 
\newcount\q 
\newcount\qq \qq=0 
 
\def\pref #1#2#3#4#5{\frenchspacing \global \advance \q by 1 
\edef#1{\the\q}{\ifnum \zz=1 { %
\item{[\the\q]}{#2} {\bf #3},{ #4.}{~#5}\medskip} \fi}} 
 
\def\bref #1#2#3#4#5{\frenchspacing \global \advance \q by 1 
\edef#1{\the\q}{\ifnum \zz=1 { %
\item{[\the\q]}{#2}, {\it #3} {(#4).}{~#5}\medskip} \fi}} 
 
\def\gref #1#2{\frenchspacing \global \advance \q by 1 
\edef#1{\the\q}{\ifnum \zz=1 { %
\item{[\the\q]}{#2}\medskip} \fi}}

\def\sref #1{~[#1]} 
 
\def\references#1{\zz=#1 
\parskip=2pt plus 1pt 
{\ifnum \zz=1 {\noindent \bf References \medskip} \fi} \q=\qq 
\pref{\bse}{E.~E.~Salpeter and H.~A.~Bethe, Phys.~Rev.}{84}{1232 (1951)}{} 
\pref{\se}{E.~E.~Salpeter, Phys.~Rev.}{87}{328 (1952)}{} 
\bref{\lieb}{E.~H.~Lieb and M.~Loss}{Analysis}{American Mathematical Society, 
New York, 1996} {The definition of the Salpeter kinetic-energy operator is 
given on p.~168.} 
\pref{\hallwnh}{R. L. Hall, W. Lucha, and F. F. Sch\"oberl, J. Math. Phys.}{43}
{1237 (2002); Erratum {\it ibid.} {\bf 44}, 2724 (2003)}{} 
\pref{\hous}{W. M. Houston, Phys.~Rev.}{47}{942 (1935)}{} 
\pref{\post}{H. R. Post, Proc. Phys. Soc. London}{66}{942 (1953)}{}
\pref{\hallaa}{R.~L.~Hall, Phys.~Rev.\ A}{51}{3499 (1995)}{}
\pref{\hallab}{R.~L.~Hall, Can.~J.~Phys.}{50}{305 (1972)}{} 
\pref{\hallac}{R.~L.~Hall, Aequ.~Math.}{8}{281 (1972)}{} 
\pref{\hallad}{R.~L.~Hall, J. Math. Phys.}{29}{990 (1988)}{} 

} 
 
\references{0} 
 
\topskip=20pt 
\trr 
\ppl{CUQM-101}\ppl{HEPHY-PUB 774/03}\ppl{UWThPh-2003-XX}\ppl{math-ph/0311032} 
\ppl{November 2003}\medskip 
\vskip 0.4 true in 
\centerline{\trbig \ptitle} 
\vskip 0.2 true in 
\centerline{\trbig \ptitlee} 
\vskip 0.4true in 
\baselineskip 12 true pt 
\centerline{\bf Richard L.~Hall$^1$, Wolfgang Lucha$^2$, and Franz 
F.~Sch\"oberl$^3$}\medskip 
\nll $^{(1)}${\sl Department of Mathematics and Statistics, Concordia 
University, 1455 de Maisonneuve Boulevard West, Montr\'eal, Qu\'ebec, Canada 
H3G 1M8} 
\nll $^{(2)}${\sl Institut f\"ur Hochenergiephysik, \"Osterreichische 
Akademie der Wissenschaften, Nikolsdorfergasse 18, A-1050 Wien, Austria} 
\nll $^{(3)}${\sl Institut f\"ur Theoretische Physik, Universit\"at Wien, 
Boltzmanngasse 5, A-1090 Wien, Austria} 
  
\nll{\sl rhall@mathstat.concordia.ca, wolfgang.lucha@oeaw.ac.at, 
franz.schoeberl@univie.ac.at} 
\bigskip\medskip 
\baselineskip = 18true pt 
 
\centerline{\bf Abstract}\medskip 
 
\noindent We study the lowest energy $E$ of a semirelativistic system of $N$ 
identical massless bosons with Hamiltonian
$$H= \sum_{i=1}^N\sqrt{\mb{p}_i^2}+\sum_{j>i=1}^N\gamma|\mb{r}_i-\mb{r}_j|^2, 
\quad\gamma>0.$$
\nl We prove
$$A\left(\gamma N^2 (N-1)^2\right)^{1\over 3}\quad\leq\quad E\quad\leq\quad B\left(\gamma N^2 (N-1)^2\right)^{1\over 3},$$
\nl where $A = 2.33810741$ and $B = \left({81\over{2\pi}}\right)^{1\over 3} = 2.3447779.$ The average of these bounds determines $E$ with an error less than $0.15\%$ for all $N \geq 2.$
 
\medskip\noindent PACS: 03.65.Ge, 03.65.Pm, 11.10.St 
\np 
One of the advantages of studying the semirelativistic ``spinless-Salpeter'' Hamiltonian\sref{\bse,\se} is
that it captures some aspects of a full relativistic treatment and at the same time allows us to
express the many-body problem in a tractable form.  The principal source of mathematical difficulty is
 the kinetic-energy operator $\sqrt{m^2 + \mb{p}^2}$, which is defined in momentum space as a multiplicative operator\sref{\lieb}, and
 becomes, via the Fourier transform, a non-local operator in configuration space.  We have earlier found energy bounds\sref{\hallwnh} for 
systems of $N$ bosons in the case $m \geq 0.$ In the nonrelativistic limit $m \rightarrow \infty$ the kinetic energy has the Schr\"odinger asymptotic form $\sqrt{m^2 + \mb{p}^2}\ \simeq\  m + {{\mb{p}^2}\over{2m}}.$ Since the Schr\"odinger many-body harmonic-oscillator problem is exactly soluble\sref{\hous-\hallaa}, we were able to derive energy bounds that are  asymptotically exact as $m\rightarrow \infty.$  The bounds were weakest in the ultrarelativistic limit $m\rightarrow 0.$ It is the purpose of this paper to present accurate bounds for this limiting case $m = 0.$
\nl The Hamiltonian for the system we study is given by 
$$H= \sum_{i=1}^N\sqrt{\mb{p}_i^2}+\sum_{j>i=1}^N\gamma|\mb{r}_i-\mb{r}_j|^2, 
\quad\gamma>0.\eqno{(1)}$$
\nl We shall prove that the lowest energy $E$ of this system satisfies the inequalities
$$A\left(\gamma N^2 (N-1)^2\right)^{1\over 3}\quad\leq\quad E\quad\leq\quad B\left(\gamma N^2 (N-1)^2\right)^{1\over 3},\eqno{(2)}$$
\nl where the coefficients $A$ and $B$ are given by
$$A = {\rm Ai}(0) = 2.33810741,\quad {\rm and}\quad B = \left({81\over{2\pi}}\right)^{1\over 3} = 2.3447779.\eqno{(3)}$$
\nl The energy $E$ of the \hi{N}{body} system is therefore determined by the average of the bounds in Eq.~(2) for all couplings $\gamma > 0,$ and all $N\geq 2,$ with an error less than $0.15\%.$
\nl In order to establish the energy bounds we must consider two fundamental symmetries: translational invariance, and boson permutation symmetry. The Hamiltonian $H$ includes the kinetic energy of the centre of mass.  Therefore we choose a set of relative coordinates so that the kinetic energy of the centre of mass can be eliminated.  The most convenient relative coordinates for our purposes are Jacobi coordinates defined in terms of the (column) vector $[\mb{r}]$ of individual-particle coordinates by an orthogonal matrix $R.$ Thus we write 
$[\rho] = R[\mb{r}].$ Since $R$ is orthogonal, the conjugate momenta $[\pi]$ are given in terms of the individual momenta $[\mb{p}]$ by the expression $[\pi] = R[\mb{p}].$ The first of the Jacobi coordinates, $\rho_1,$ is proportional to the centre-of-mass variable, so that the elements of the first row are all equal to $1/\sqrt{N}.$ The other two coordinates which we shall need to refer to specially are $\rho_2$ and $\mb{p}_N$ which are given explicitly in terms of the `other set' by
$$\rho_2 = {{\mb{r}_1 - \mb{r}_2}\over{\sqrt{2}}},\quad \mb{p}_N= 
{{1}\over{\sqrt{N}}}\pi_1-\sqrt{{{N-1}\over N}}\pi_N.\eqno{(4)}$$ 
\nl The expression of the boson permutation-symmetry constraint in terms of Jacobi coordinates can be a source of complication. But we do not need to face this difficulty here: we simply exploit the `reducing power' of the necessary boson symmetry to relate the \hi{N}{body} problem to a scaled two-body problem.  Let us assume that $\Psi$ is a normalized boson wave function of the $N-1$ relative coordinates $\{\rho_i\}_{i = 2}^{N}.$ By Lemma~(1) established in Ref.\sref{\hallwnh},
we know that an operator acting on $\Psi$ with leading term $\pi_1,$ may be replaced by zero, even when the term is inside the kinetic-energy square root.  This will be important later.  We shall also use another important relation\sref{\hallwnh, Eq.~(2.5)}, namely 
$$\left(\Psi,\rho_i^2\Psi\right)=\left(\Psi,\rho_2^2\Psi\right),\quad 2\le i\le 
N.\eqno{(5)}$$
\nl An arbitrary boson wave function is not necessarily symmetric in the $\{\rho_i\}_{i = 2}^{N},$ but Eq.~(5) {\it is} generally true, and is very useful.  A special boson wave function which certainly is symmetric in the $\{\rho_i\}_{i = 2}^{N}$ is the Gaussian function $\Psi_g$ which also has another unique\sref{\hallab,\hallac} and useful property,
namely it factors into single-variable Gaussians $\psi_g$ as follows:
$$\Psi_g(\rho_2,\rho_3,\dots,\rho_N) 
=\prod_{i = 2}^{N}\psi_g(\rho_i),\quad \psi_g(r) = \left({{a}\over{\pi}}\right)^{3\over 4}\exp\left(-{{a r^2}\over 2} \right),\quad a > 0.\eqno{(6)}$$
\nl That $\Psi_g$ has the correct boson symmetry follows immediately from the following identity
valid for Jacobi relative coordinates
$$N\sum_{i = 2}^{N} \rho_i^2 = \sum_{1=i<j}^{N}|\mb{r}_i-\mb{r}_j|^2.\eqno{(7)}$$
\nl In momentum space the Gaussian transforms to a Gaussian by the three-dimensional Fourier transform ${\cal F}_3$ as follows:
$$\phi_g = {\cal F}_3(\psi_g),\quad \phi_g(k) = \left({1\over{a\pi}}\right)^{3\over 4} \exp\left(-{{k^2}\over{2a}}\right).\eqno{(8)}$$\medskip

The lower energy bound is found by the following argument.  We suppose that $\Psi$ is the
exact ground-state wave function for the \hi{N}{body} system corresponding to energy $E$ and, using the necessary boson symmetry, we write
$$E = (\Psi, H\Psi) = \left(\Psi, \left\{N\sqrt{\mb{p}_N^2} + {N\choose 2}\gamma|\mb{r}_1-\mb{r}_2|^2\right\}\Psi\right).$$
\nl By employing (4) and (5) in succession, and noting that the lemma allows us to remove the operator $\pi_1$ from the square root, we arrive at the relation
$$E = (\Psi, H\Psi) = \left(\Psi, \left\{\alpha^{1\over 2}\sqrt{\pi_N^2} + \alpha\gamma\rho_N^2\right\}\Psi\right),\quad  {\rm where}\quad \alpha = N(N-1).$$
\nl Thus the \hi{N}{body} energy $E$ is bounded below by the lowest energy ${\cal E}^L$ of the one-body Hamiltonian
$${\cal H} = \alpha^{1\over 2}\sqrt{\mb{p}^2} + \alpha\gamma\mb{r}^2 \quad{{{\cal F}_3}\atop{\longrightarrow}} \quad\alpha^{1\over 2}r + \alpha\gamma\mb{p}^2.$$
\nl But ${\cal F}_3({\cal H})$ is the Hamiltonian for the Schr\"odinger problem of a single particle moving in a linear potential $r.$ Thus we find
$$E \geq {\cal E}^L = A(\gamma\alpha^2)^{1\over 3},\quad {\rm where}\quad A = {\rm Ai}(0) \approx 2.33810741$$
\nl is the first zero of Airy's function, and is also exactly the bottom of the spectrum of $\mb{p}^2 + r$ in three dimensions.  This establishes the lower energy bound.\medskip

The upper bound is found by means of the  Gaussian `trial' function $\Psi_g$ discussed above.  We have
$$E \leq {\cal E}^U(a) = (\Psi_g,H\Psi_g) = \alpha^{1\over 2}(\phi_g,k\phi_g) + \alpha\gamma(\psi_g,r^2\psi_g).$$
\nl That is to say
$${\cal E}^U(a) = \left({{4\alpha a}\over{\pi}}\right)^{1\over 2} + {{3\alpha\gamma}\over{2a}}.$$
\nl By minimizing with respect to the variational parameter $a > 0$ we obtain
$$ E \leq {\cal E}^U = B(\gamma\alpha^2)^{1\over 3},\quad {\rm where}\quad B = \left({{81}\over{2\pi}}\right)^{1\over 3}\approx 2.3447779.$$
\nl This result establishes the upper bound.  It is perhaps tempting to try to improve the upper bound by the use of a more flexible trial function.  However, it is not trivially easy to accomplish this, and to keep the calculation and result simple, since we must use a translation-invariant boson function.  \medskip

It is interesting that there is a relationship between the problem discussed in this paper and the corresponding 
 Schr\"odinger problem with a linear potential and $m > 0.$  To be more precise, 
if we consider the nonrelativistic problem with Hamiltonian $\tilde{H}$ given by
$$\tilde{H} = \sum_{i=1}^N{{\mb{p}_i^2}\over{2m}}+\sum_{j>i=1}^N\lambda|\mb{r}_i-\mb{r}_j|, 
\quad\lambda>0,\eqno{(9)}$$
\nl then we have shown\sref{\hallad,~Eq.~(4.16)} that the lowest energy $\tilde{E}$ of $\tilde{H}$ is bounded by the inequalities
$$A \left(N^{2}(N-1)^{3}{{\lambda^{2}}\over{4m}}\right)^{1\over 3}\quad\leq
\quad \tilde{E}\quad\leq\quad B \left(N^{2}(N-1)^{3}{{\lambda^{2}}\over{4m}}\right)^{1\over 3},\eqno{(10)}$$
\nl where the coefficients $A$ and $B$ are exactly as given in Eq.~(3) above. The two problems are brought more nearly into `coincidence' if we set $2m = 1.$ The remaining differences, involving $N$ and $\lambda$, can then be understood if one notes that the ${N\choose 2}$ factor must be associated with the `potential terms' in each case, and, similarly, the $\lambda$ and $\gamma$ couplings must be made to `correspond' by scaling, as the Fourier transformation, which relates the two systems, is applied.  With the aid of such arguments,  the `almost equivalence' of the problems could eventually be used formally to extract our main result.  However, we present these considerations here merely as a confirmation of our results; we prefer to develop more widely applicable direct approaches for the semirelativistic many-body problem itself, valid for all $m \geq 0.$

\title{Acknowledgements} 
 
 Partial financial support of this work under Grant No. GP3438 from the Natural 
Sciences and Engineering Research Council of Canada, and the hospitality of the 
Institute for High Energy Physics of the
Austrian Academy of Sciences in Vienna, is gratefully acknowledged by one of us
 [RLH].\medskip 
 
\np 
\references{1} 

\end